%% file: ospDisplayWall.tex
\documentclass[prepring]{vgtc}
\usepackage{relsize}

\usepackage[T1]{fontenc}
\usepackage[utf8]{inputenc}
\usepackage{microtype}                 
\PassOptionsToPackage{warn}{textcomp}  
\usepackage{textcomp}                  
\usepackage{mathptmx}                  
\usepackage{times}                     
\usepackage{cite}                      
\usepackage{tabu}                      
\usepackage{booktabs}                  
\usepackage[dvipsnames]{xcolor}
\usepackage{soul}
\usepackage{listings}
\usepackage{url}
\usepackage[noend]{algpseudocode}

  \usepackage[pdftex]{graphicx}
  \DeclareGraphicsExtensions{{.png},{.pdf},{.jpg},{jpeg}}

\usepackage{color}

\onlineid{0}

\vgtccategory{Research}

\vgtcinsertpkg

\usepackage[labelfont=sf]{subcaption}
\usepackage{cite}
\usepackage{cleveref}

\definecolor{dkgreen}{rgb}{0,0.6,0}
\definecolor{gray}{rgb}{0.5,0.5,0.5}
\definecolor{mauve}{rgb}{0.58,0,0.82}

\definecolor{dkgreen}{rgb}{0,0.6,0}
\definecolor{dkblue}{rgb}{0,0,0.6}
\definecolor{gray}{rgb}{0.5,0.5,0.5}
\definecolor{mauve}{rgb}{0.58,0,0.82}

\definecolor{commentgreen}{RGB}{2,112,10}
\definecolor{eminence}{RGB}{108,48,130}
\definecolor{weborange}{RGB}{255,165,0}
\definecolor{frenchplum}{RGB}{129,20,83}

\title{A Virtual Frame Buffer Abstraction for Parallel Rendering of Large Tiled Display Walls}

\author{
  Mengjiao Han\thanks{e-mail: mengjiao@sci.utah.edu}$^*$
  \and
  Ingo Wald$^{\ddag}$
  \and
  Will Usher$^{*, \dagger }$
  \and
  Nate Morrical$^*$
  \and 
  Aaron Knoll$^\dagger$
  \and
  Valerio Pascucci$^*$
  \and
  Chris R. Johnson$^*$
  \vspace{-3em}
}

\affiliation{
	\scriptsize
  $^*$SCI Institute, University of Utah
  \quad\quad
  $^\ddag$NVIDIA Corp
  \quad\quad
  $^\dagger$Intel Corp.
}

\teaser{
  \vspace{-2em}
	\centering
  \includegraphics[width=0.495\linewidth]{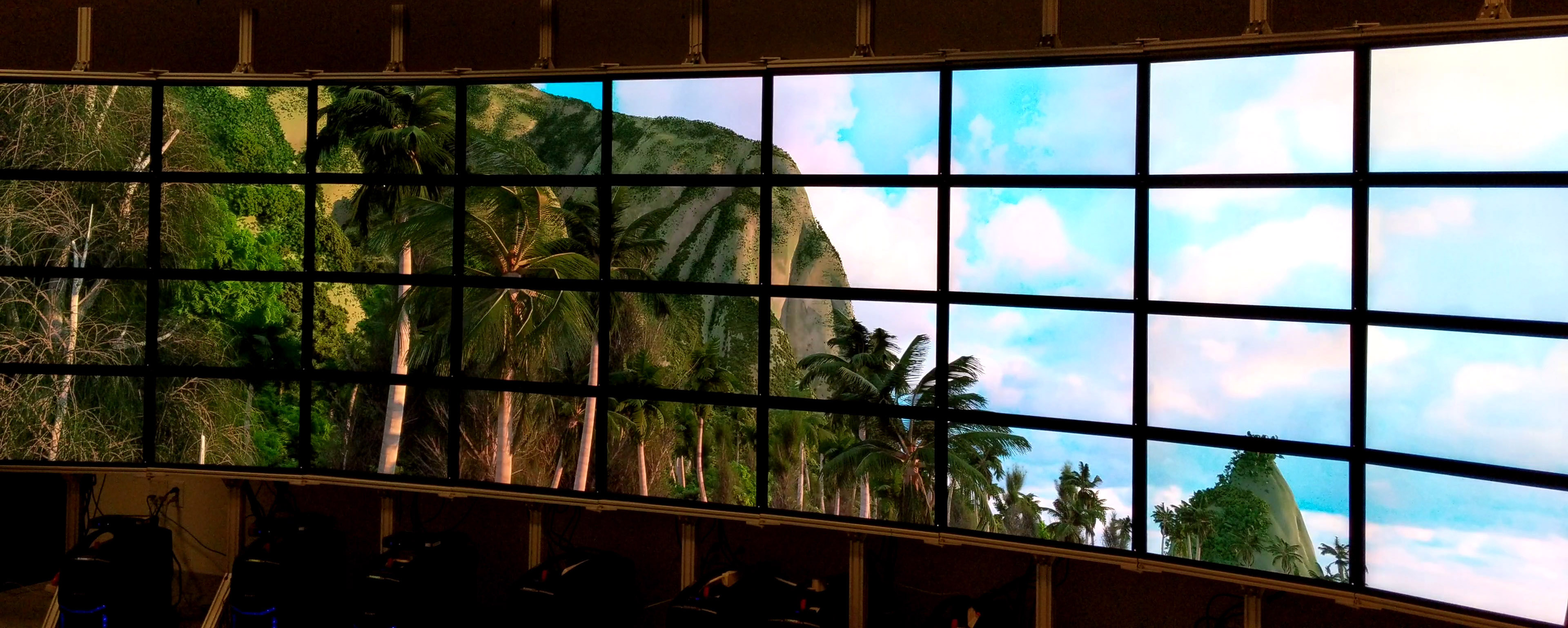}
  \includegraphics[width=0.495\linewidth]{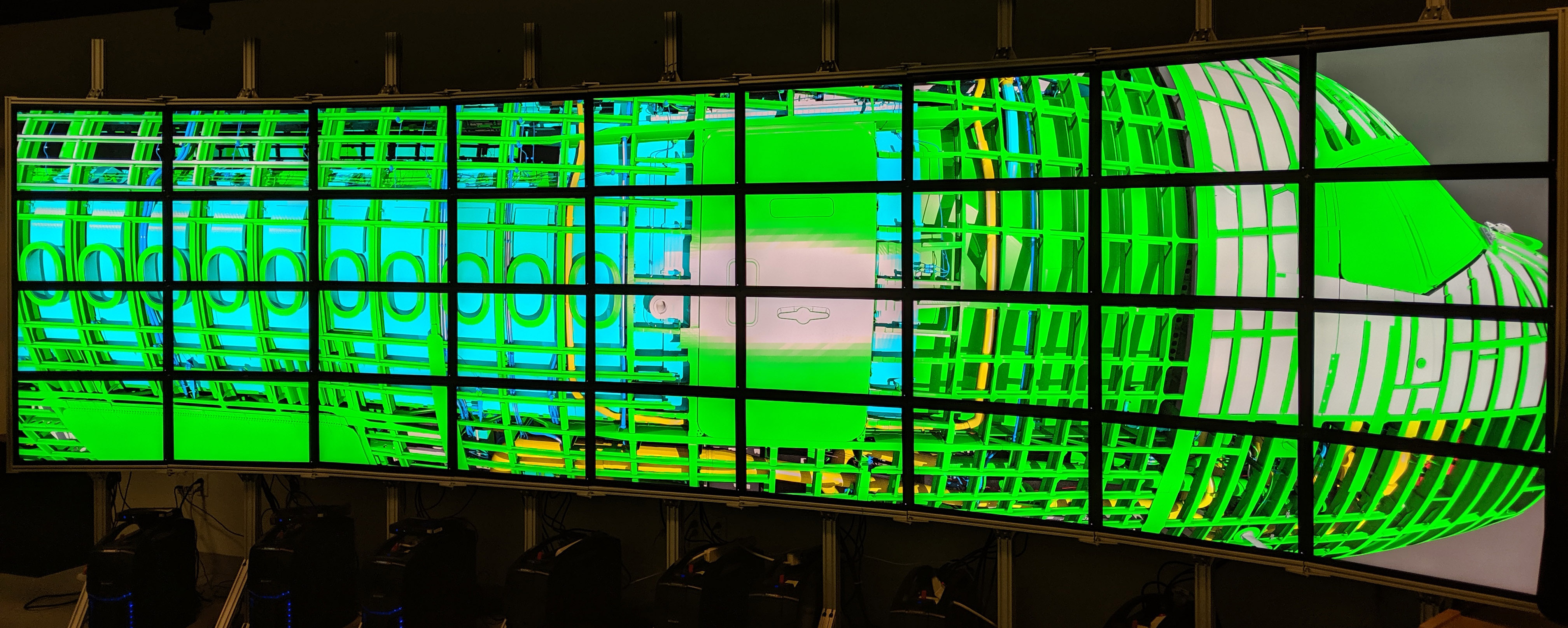}
	\vspace{-1em}
	\caption{\label{fig:teaser}%
	Left: The Disney Moana Island~\cite{moana_whitepaper} rendered remotely
	with OSPRay's path tracer at full detail using 128 Skylake Xeon (SKX)
	nodes on Stampede2 and streamed to the 132Mpixel \emph{POWERwall} display wall,
	averages 0.2-1.2~FPS.
	Right: The Boeing 777 model, consisting of 349M triangles, rendered
	remotely with OSPRay's scivis renderer using 64 Intel Xeon Phi Knight's Landing
	nodes on Stampede2 and streamed to the POWERwall, averages 6-7~FPS.}
	\vspace{-0.5em}
}
\abstract{%
  We present \emph{dw2}, a flexible and easy-to-use software infrastructure for
  interactive rendering of large tiled display walls.
  Our library represents the tiled display wall as a single virtual screen
  through a display ``service'',
  which renderers connect to and send image tiles to be displayed,
  either from an on-site or remote cluster.
  The display service can be easily configured to support a range of typical
  network and display hardware configurations; the client
  library provides a straightforward interface for easy integration
  into existing renderers.
  We evaluate the performance of our display wall service in different
  configurations using a CPU and GPU ray tracer, in both on-site and remote rendering
  scenarios using multiple display walls.
}

\CCScatlist{
  Tiled Display Walls;
  Distributed Display Frameworks
  \vspace{-2mm}
}

\begin{document}

\maketitle


\input{introduction.tex}

\input{background.tex}

\input{framework.tex}
\input{ospray-integration.tex}

\input{results.tex}

\input{summary.tex}

\input{acks.tex}







\bibliographystyle{abbrv-doi}
\bibliography{ospDisplayWall}

\end{document}

%% file: introduction.tex
\section{Introduction}

Tiled displays are important communication tools in modern visualization
facilities. They are beneficial to visualization in many ways: 
displaying the features of data at a large scale increases the user's sense of 
immersion, better conveys a sense of scale (e.g., when viewing an entire car or
airplane), and the high resolution provided is valuable when visualizing highly detailed
datasets (\Cref{fig:teaser}).
Perhaps most importantly, tiled displays are powerful communication 
tools and can engage a large group of collaborators simultaneously.

A number of high-end visualization facilities feature tiled displays, using either multiprojector systems, CAVEs~\cite{cruz1992cave, reda2013visualizing}, 
or multiple high-resolution LED panels---such as TACC's 189~MPixel \emph{Rattler}
display wall and 328~MPixel \emph{Stallion}, 
NASA's 245~MPixel HyperWall~2, or SUNY StonyBrook's RealityDeck. 
Unfortunately, the exact requirements, configurations, and software stacks for such tiled display walls vary greatly across systems,
and thus there is no easy or standardized way to use them~\cite{chung2013survey}.
Visualization centers often build their own proprietary software for driving such walls,
requiring system-specific modifications to each software package to use the wall. 
Typical software set-ups often assume that each display node will render the pixels for its attached
display~\cite{chromium,paul2008chromium,eilemann2009equalizer}. Rendering on the display nodes is sufficient for moderately
sized datasets but not for large-scale ones. 
To support large data, systems typically render on an HPC cluster and stream the image back to the display wall.

DisplayCluster~\cite{displayCluster} and SAGE2~\cite{sage2} are two general 
and widely used streaming frameworks 
for tiled display walls that can support local and remote collaborations with multiple devices, such as kinect, 
touch overlays, or smart phones/tablets. One disadvantage is that communication with the display wall must
be performed through a master node. The master node, therefore, must be powerful enough to process and stream 
all the pixels for the entire display wall to avoid becoming a bottleneck. 
DisplayCluster is used for scientific visualization as it supports 
distributed visualization applications using IceT~\cite{moreland_icet}.
However, IceT, a sort-last compositing framework, is less well suited for 
large tile-based ray tracing applications~\cite{usher2019scalable}.

In this paper, we describe a lightweight open-source framework for driving tiled display walls from a 
single node or distributed renderer.
In our framework, the display wall is treated as 
a single virtual frame buffer managed through a display service.
Our framework supports both dispatcher and direct communication modes
between the rendering clients and display service to support typical
network configurations.
The direct mode can relieve network congestion and the bottleneck on the master node, which makes it possible to 
use low-cost hardware for display walls, e.g., the Intel NUC mini PCs~\cite{intel}. 
Moreover, our framework can easily be used by both CPU and GPU renderers for portability.
We demonstrate integration of our library
into OSPRay~\cite{wald_ospray_2017} and a prototype GPU raycaster~\cite{wald_rtx_2019}
for interactive rendering on typical tiled display walls and low-cost display walls.
Our contribution are:

\begin{itemize}
    \item We present a lightweight open-source framework for driving tiled display walls that can be integrated into 
    CPU and GPU renderers;
    \vspace{-0.5em}
    \item The framework can transparently operate in the dispatcher or direct mode to support typical network configurations;
    \vspace{-0.5em}
    \item We demonstrate this framework for use in deploying low-cost alternatives for display walls.
\end{itemize}



%% file: background.tex
\section{Related Work}

\subsection{Cluster-Based Tiled Display Walls}

A large number of supercomputing centers now use
a tiled display wall for some of their high-end
visualizations. These systems come in a mix of configurations, 
in terms of the display layout, hardware used to drive the displays,
and network connectivity to local and remote HPC resources.
For example, TACC's Stallion and Rattler systems and NASA's Hyperwall~2
use a single node per display; however, the POWERwall
at the Scientific Computing and Imaging (SCI) Institute uses one node
per column of four displays. Each node on the POWERwall is directly
accessible over the network, and on Hyperwall~2, each node
is connected directly to Pleiades. However, on Stallion and Rattler, 
the display nodes are not externally visible and must be accessed through a head node.
We refer to the survey by Chung et al.~\cite{chung2013survey}
for a more in-depth discussion of tiled display wall frameworks.

\subsection{GPU Rendering on Tiled Displays}

Parallel rendering on a cluster based on OpenGL is a common solution 
for driving tiled displays. Eilemann et al.~\cite{eilemann2012parallel}
presented an experimental analysis of  the
important factors for performance of parallel rendering on multi-GPU clusters.
The basic approach for OpenGL-based applications is to run an
instance of the application on each node, with a master node
used to broadcast user interactions to the display nodes.
The Chromium project~\cite{chromium}, an automatic method for such approaches, 
intercepts the application's OpenGL command stream and broadcasts it to the worker nodes. 
The Chromium Renderserver~\cite{paul2008chromium} also supports 
the distributed-memory parallel rendering using Chromium.
However, it is inherently limited by the available processing
power on the display nodes, requiring powerful on-site hardware. 

An alternative to having each node render the pixels for
its display is to use a compositing or pixel routing framework
that can route pixels from the render nodes to the corresponding
display node. One of the first methods using such an approach 
was described by Moreland et al.~\cite{moreland_display_wall},
who used a sort-last compositing scheme for rendering to tiled
display walls. The same general approach is now available in
IceT~\cite{moreland_icet}, where users can specify
a number of output windows and provide a callback to render
specific frusta for the displays. Equalizer~\cite{eilemann2009equalizer}, 
introduced by Eilemann et al., supports scalable parallel rendering and 
can distributed rendering works directly to worker nodes.
However, Chromium and Equalizer are all specific to OpenGL, 
and IceT is less applicable to tile-based ray tracers. 
Moreover, these frameworks impose the rendering work distribution 
on the application, and are not suited to applications that perform more complex
load balancing.

\subsection[Distributed Display Frameworks]{Distributed Display Frameworks}

A work similar to our own for driving tiled display walls 
was proposed by Johnson et al. in the ``DisplayCluster'' framework~\cite{displayCluster}.
Similar to our proposed framework, DisplayCluster makes a clear
distinction between a display wall ``service'', which receives pixels
and presents them on the appropriate displays, and client applications,
which produce these pixels and send them to the service.
DisplayCluster assumes that the display nodes are connected over
a high-bandwidth network, but that they are not visible to the external
network and must be accessed through a head node.
The head node communicates with clients over TCP and 
broadcasts the received pixel data to the display nodes
over the Message Passing Interface (MPI)~\cite{gabriel2004open}.
The display nodes then decompress the pixel data and discard portions
of the received image that are outside their display region.
DisplayCluster has found wider use in the communities (e.g., by the Blue Brain Project),
and has been used for displaying interactive rendering from Stampede on Stallion~\cite{knoll2013rtv}.


SAGE2~\cite{sage2} is another popular windowing environment for tiled displays,
designed for collaborative workspaces on tiled display walls.
OmegaLib~\cite{omegalib} is designed for similar use cases, with
a focus on stereo tiled display environments.
DisplayCluster, SAGE2, and OmegaLib
support displaying multiple applications on
the wall simultaneously, each streaming to its own virtual window,
which can be repositioned using the library.
These libraries are more similar to full-featured
window managers, whereas, in contrast, we aim to provide 
a simple lightweight framebuffer abstraction that can be rendered 
to by a single application.


Biedert et al.~\cite{biedert2018hardware} recently demonstrated 
a parallel compositing framework for streaming from multiple asynchronously running image sources, leveraging
GPU video compression and decompression hardware. They achieve consistently real-time rates compositing
hundreds of sources into a single 4K or tiled video stream.
However, their system requires GPU-accelerated video encoding
and does not consider a synchronized full-resolution frame rate 
across the displays.


%% file: framework.tex
\section{Framework Overview}

\begin{figure}
    \centering
	\includegraphics[width=0.8\linewidth, height=8cm]{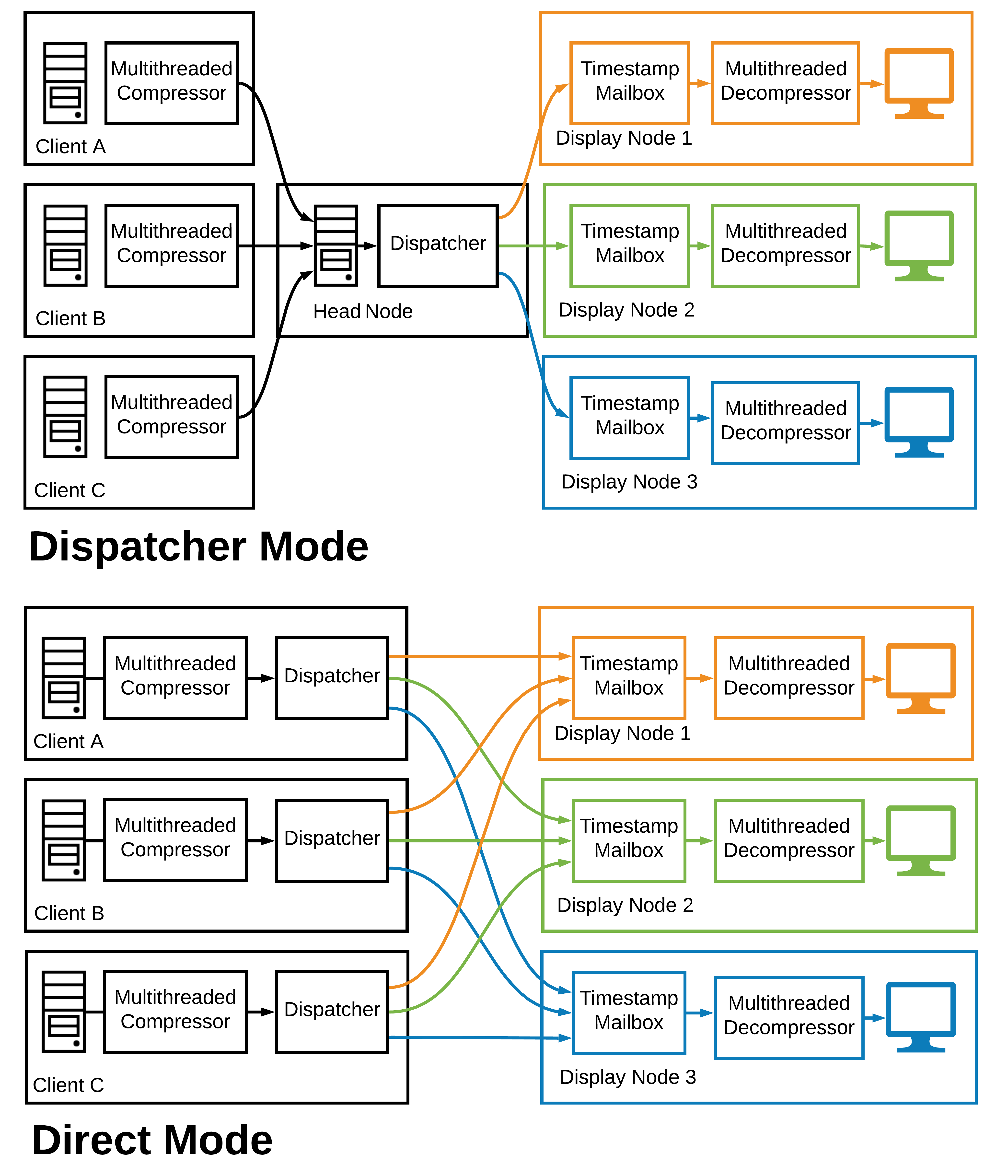}
  \vspace{-1em}
	\caption{\label{fig:framework}%
	An overview of \emph{dw2} in the dispatcher and direct communication
	modes. The best mode for the system's network architecture
	can be used as needed, without modifying the rendering client code.}
  \vspace{-1.5em}
\end{figure}

Our framework, \emph{dw2}, is split into
an MPI parallel display service that manages the mapping
from the single virtual framebuffer to the physical tiled display wall (\Cref{sec:service}) 
and a client library used to connect renderers to the display service (\Cref{sec:client}).
The client can be a single program, an MPI-parallel program,
or a distributed application with some custom parallel communication setup.
To allow for different configurations of the clients and displays, e.g., rendering on the display nodes,
on a different on-site cluster, or on a remote cluster, we use TCP sockets to communicate
between the clients and displays through a socket group abstraction. 
We provide pseudo-code of how the display service cooperates with 
the rendering clients in the supplementary material. 
Source code and detailed instructions can be found in the project repository: \url{https://github.com/MengjiaoH/display_wall}.

\subsection{Display Service}
\label{subset:service}
\label{sec:service}


The display service supports two modes: a \emph{dispatcher mode},
where a central dispatch node manages routing of tiles to the display nodes,
and a \emph{direct mode}, where clients send tiles directly to the display nodes
(see \Cref{fig:framework}).
The latter mode can achieve better network utilization and performance, 
but on some systems the display nodes are not visible to the outside
network for security reasons and must be accessed via a single externally visible node.

The display service is run using MPI, with one process launched per display on
each node. In the dispatcher mode, an additional process is needed, and rank 0 is used as
the dispatcher on the head node.
At start-up the service is passed information about the windows to open on each node,
their location on the wall, and the bezel size, to provide a single continuous image across the displays.



In both the dispatcher and direct modes, rank 0 acts as the
information server for the wall. Clients connect to the service through
this rank and receive information about the display wall's size and configuration.
In the dispatcher mode, all clients connect to the dispatcher through
a socket group. In the direct mode, clients are sent back host name
and port information for each display that they then connect to directly.
Each display process returns its size and location in the display wall
to allow each client to perform tile routing locally.
Each tile consists of an uncompressed header specifying its size and location,
along with the JPEG compressed image data.



On both the dispatcher and the display processes, multiple threads
are used for receiving and sending data, and for decompressing tiles on the
displays.
Communication between threads is managed by \emph{timestamped mailboxes},
which are locking producer-consumer queues that can be optionally filtered to return only
messages for the current frame.
Each socket group is managed by a pair of threads, one that takes outgoing
messages from the mailbox and sends them, and another that places received messages
into an incoming mailbox.
In the dispatcher mode, the dispatcher receives tiles, reads
their header, and routes them to the display processes they cover via MPI.
In the direct mode, each client tracks the individual display information
received above and runs its own dispatcher to route tiles directly to
the displays via the socket group.
Each display places incoming messages into a timestamped mailbox.
A set of decompression threads take tiles for the current frame from the mailbox,
decompress them, and writes them to the framebuffer.
Once all pixels in the virtual
framebuffer have been written, the frame is complete.


After the frame is complete, process 0 sends a
token back to the clients to begin rendering the next frame.
This synchronization prevents the renderer from running faster than
the displays, and thus a buildup of buffered tiles, causing them to run out of memory.
However, it causes a delay on how soon the renderer can start on the next frame.
To alleviate this delay, users can configure the number of frames that can
be in flight at once, allowing the renderer to begin the next frame immediately to
buffer some number of frames.
If the renderer and displays run at similar
speeds, this approach will significantly reduce the effect of latency.

\subsection[Rendering with the Client Library]{Rendering with the Client Library}
\label{sec:client}


The client library provides a small C API to allow
for easy integration into a range of rendering applications (also see supplemental materials).
Clients first query the size of the virtual framebuffer from the display
service using \texttt{dw2\_query\_info}, 
after which they connect to the service to set up a socket group.
Depending on the mode used by the display service, the library will either
connect to the dispatcher or to each individual display.
Connections are established using socket groups, where each client sends a token returned with
the initial information query and its number of peers, allowing the display process to track when all clients
have been connected.
All clients then call \texttt{dw2\_begin\_frame}, which returns when the display service
is ready to receive the next frame. 
The client can divide the image into tiles as it sees fit to distribute the rendering workload.
After a tile is rendered, the client calls \texttt{dw2\_send\_rgba} to send it to the display service.
The tile is then compressed and sent to the dispatcher or the overlapped displays by the library.
The client library also leverages multiple threads for compression and networking, in the
same manner as the display processes.

%% file: ospray-integration.tex
\section{OSPRay Integration}
\label{sec:render_ospray}
We integrate our client library into OSPRay (version 1.8) through a pixel operation
that reroutes tiles to the display wall.
Pixel operations in OSPRay are per-tile postprocessing operations
that can be used in local and MPI-parallel rendering through OSPRay's
Distributed FrameBuffer~\cite{usher2019scalable}.
After querying the display wall's dimensions, we create a single large framebuffer
with the display wall's size and attach our pixel operation it.
The framebuffer is created with the
\texttt{OSP\_FB\_NONE} color format, indicating that no final pixels
should be stored. By sending the tiles in the pixel operation
and creating a \texttt{NONE} format framebuffer, we can send tiles directly
from the node that rendered them and skip aggregating
final pixels to the master process entirely.

\section{GPU Raycaster Integration}
\label{sec:render_gpu}
The prototype GPU raycaster~\cite{wald_rtx_2019} uses OptiX~\cite{optix} (version 6.5) 
for rendering on a single node equipped with one GTX 1070 GPU. To allow rendering to large-scale display walls,
we extend the renderer with an image-parallel MPI mode that divides the image
into tiles and assigns them round-robin to the processes.
On each rank, we create a tiled framebuffer containing the tiles it owns
and render them using the prototype's existing renderer code.
After the tiles are rendered, each rank passes its tiles to \emph{dw2}
to be sent to the displays.
To achieve interactive performance at high resolution, we also extend the
rendered with a screen-space subsampling strategy.


%% file: results.tex
\section{Experiments and Results}
\label{sec:results}

We evaluate the performance of \emph{dw2} in on-site and remote streaming
rendering scenarios to study the performance of the dispatcher and
direct modes, the impact of compression and the client's chosen tile size on performance,
and scalability with the number of clients and displays in \Cref{sec:dw2_perf}. 
We demonstrate interactive rendering use cases of \emph{dw2}
on a range of datasets in \Cref{sec:use_cases} using OSPRay and the GPU renderer.

We conduct our evaluation on three tiled display wall systems:
the POWERwall and NUCwall at SCI and Rattler at TACC. The POWERwall has
a $9\times4$ grid of $2560\times1440$ monitors (132Mpixel),
with each column of four monitors driven by one node, along with an
optional head node; each node has an i7-4770K CPU.
The NUCwall has a $3\times4$ grid of $2560\times1440$ monitors (44Mpixel),
with each column of four monitors driven by an Intel NUC (i7-8809G CPU).
We run on a subset of TACC's Rattler, a $3\times3$ grid of 4K monitors (74Mpixel), 
with each display driven by a node with an Intel Xeon E5-2630 v3 CPU.
The POWERwall and NUCwall use the same network configuration,
where each node has a 1Gbps ethernet connection and is accessible externally.
Rattler's display nodes are not accessible externally and are connected
to a head node using a 1Gbps network, with a 1Gbps connection from the head node to Stampede2.

\subsection{dw2 Performance Evaluation}
\label{sec:dw2_perf}
\begin{figure}[t]
	\centering
	\begin{subfigure}{0.48\columnwidth}
		\includegraphics[width=\textwidth]{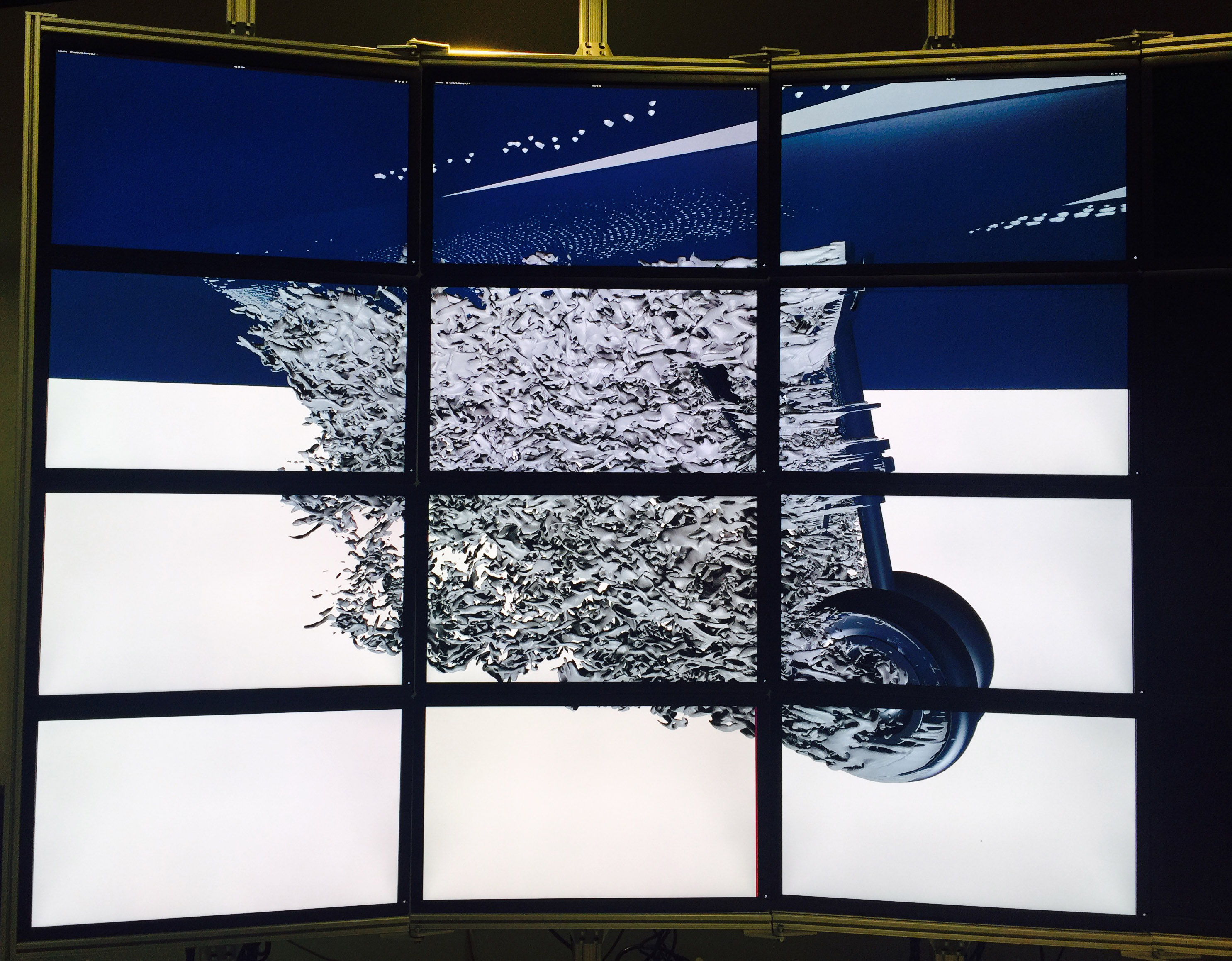}
		\vspace{-1.5em}
		\caption{\label{fig:nucs}%
		Landing Gear, on the NUCwall}
	\end{subfigure}
	\begin{subfigure}{0.48\columnwidth}
		\includegraphics[width=\textwidth]{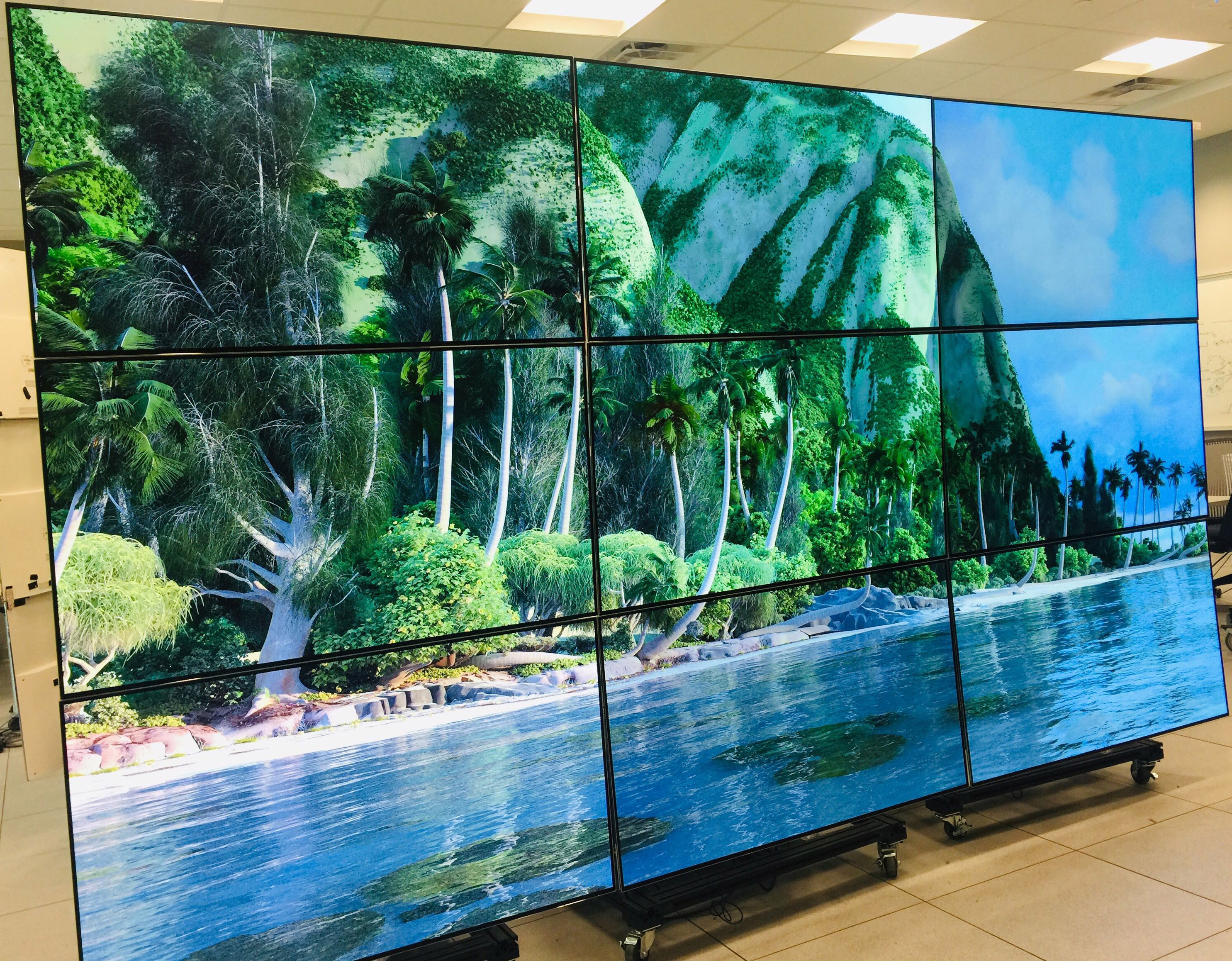}
		\vspace{-1.5em}
		\caption{\label{fig:rattler_stampede2_moana}%
		The Moana Island, on Rattler.}
	\end{subfigure}
	  \vspace{-1em}
	  \caption{\label{fig:images_example}%
	  The test images and use cases of (a) the landing gear remote rendering to the low-cost NUCwall
      and (b) the Moana Island Scene on-site rendering to Rattler.
      \vspace{-1.5em}}
  \end{figure}
  
To isolate the performance impacts of the different configurations of \emph{dw2}
from the renderer's performance, our benchmarks are run using pre-rendered images created using OSPRay.
These images are representative of typical visualization and rendering use cases on
display walls, and they vary in how easily they can be compressed. The Landing Gear contains a complex isosurface with a large 
amount of background and compresses well, and the Moana Island Scene contains high-detail geometry and 
textures and is challenging to compress (see~\Cref{fig:images_example}).
Additionally, we benchmark on a generated image with varying colors
within each tile to provide a synthetic benchmark case that is difficult to compress.
For on-site client benchmarks, we use a local cluster with eight Intel Xeon Phi KNL 7250 processors;
remote rendering benchmarks use eight KNL 7250 nodes on Stampede2.

\begin{figure}[t]
	\centering
	\vspace{-1em}
	\begin{subfigure}{\linewidth}
		\includegraphics[width=\linewidth, height=2.5cm]{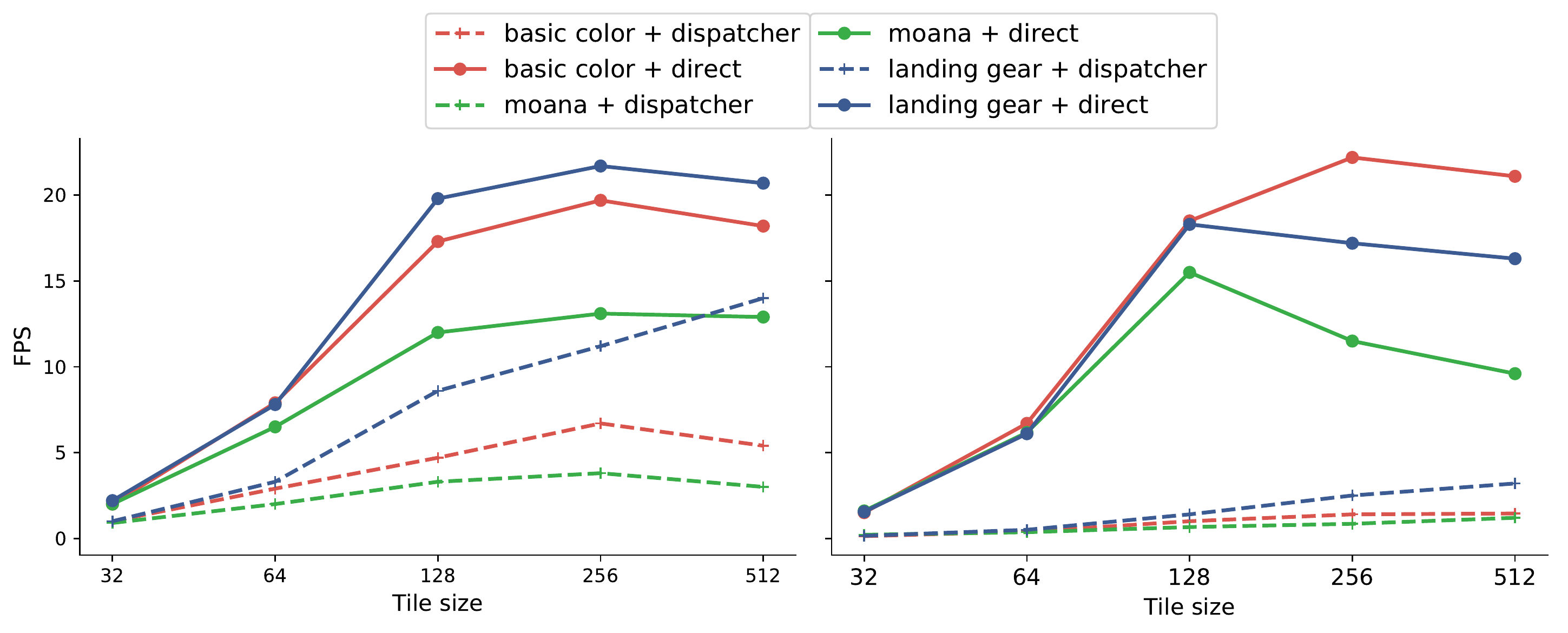}
	\vspace{-2em}
		\caption{\label{fig:tile-size}%
		Tile Size.}
	\end{subfigure}
	\begin{subfigure}{\linewidth}
	\includegraphics[width=\linewidth, height=2.5cm]{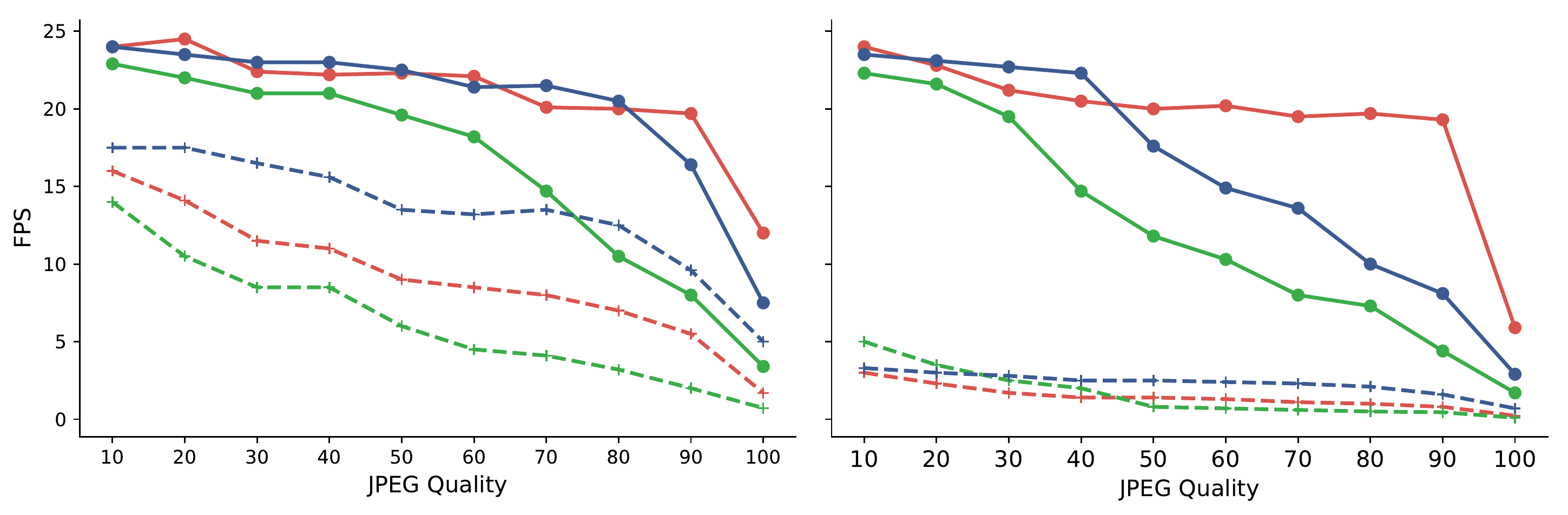}
	\vspace{-2em}
		\caption{\label{fig:compression}%
			JPEG Quality.}
	\end{subfigure}
	\vspace{-1em}
	\caption{\label{fig:param_study}%
	The performance impact of different tile sizes and JPEG quality settings
	in both modes on the POWERwall. Left: Clients run on-site on an eight-node KNL cluster. Right: Clients run remotely on eight KNL nodes on Stampede2.}
	\vspace{-1.25em}
\end{figure}


\begin{figure}[t]
	\centering
	\begin{subfigure}{\linewidth}
		\includegraphics[width=\linewidth, height=3cm]{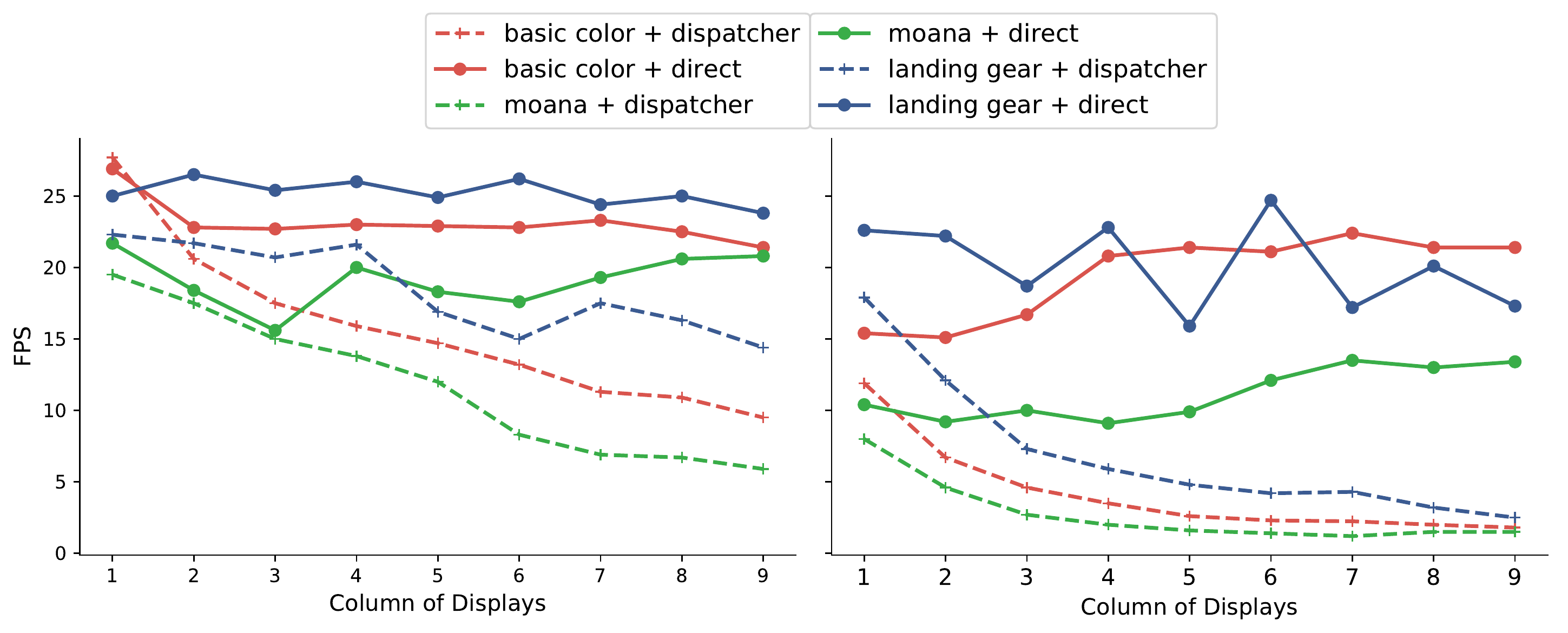}
	\vspace{-1.5em}
		\caption{\label{fig:display_scaling}%
		Scaling with the number of displays with 8 clients, each column has 4 display processes.}
	\end{subfigure}
	\begin{subfigure}{\linewidth}
		\includegraphics[width=\linewidth, height=2.5cm]{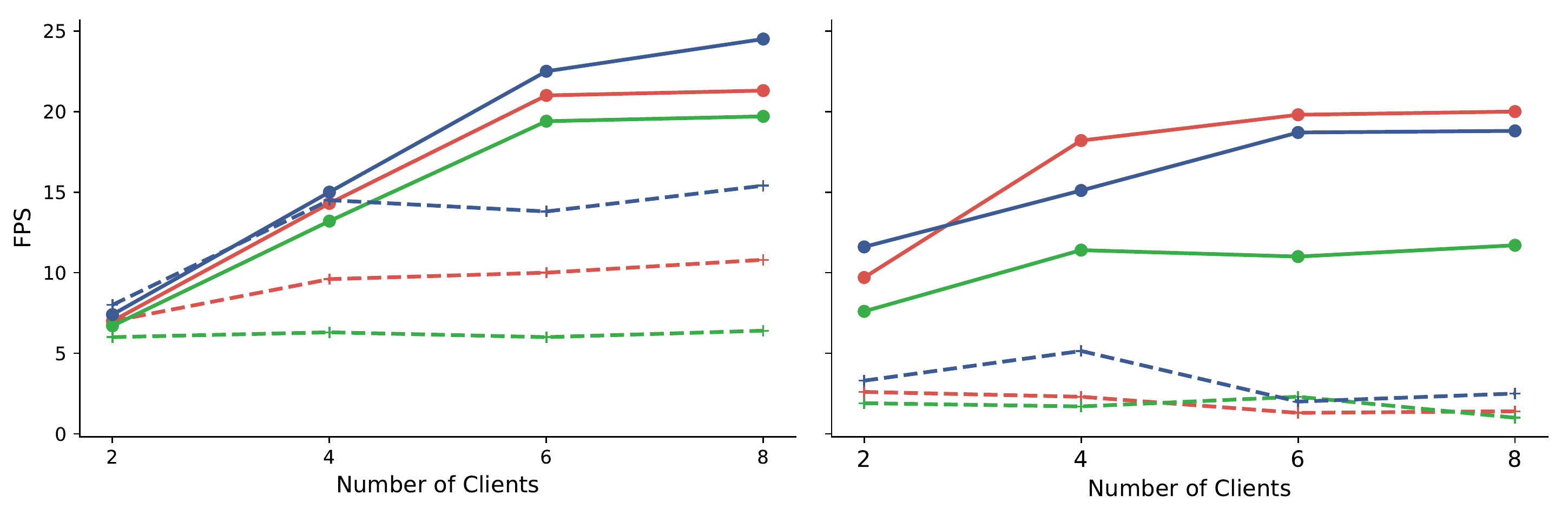}
		\vspace{-1.5em}
		\caption{\label{fig:client_scaling}%
			Scaling with the number of clients sending to all 9 display columns.}
	\end{subfigure}
	\vspace{-1em}
	\caption{\label{fig:scaling}%
	Scalability studies on the POWERwall.
	Left: Clients run on-site on an eight-node KNL cluster.
	Right: Clients run remotely on eight KNL nodes on Stampede2.}
	\vspace{-1.25em}
\end{figure}

\begin{figure}[t]
	\centering
	   \vspace{-1em}
	\includegraphics[width=0.8\columnwidth]{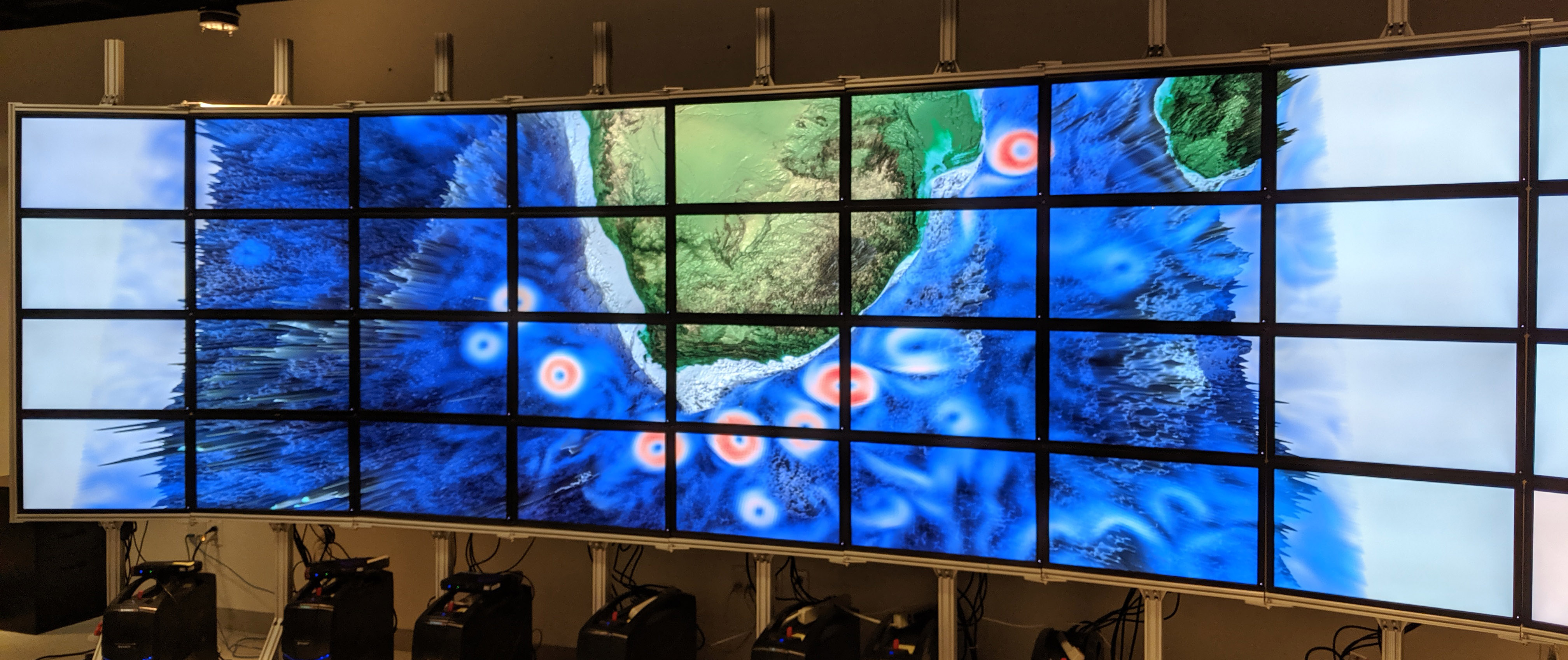}
	   \vspace{-0.5em}
	  \caption{\label{fig:sci_tetty_rtx}%
	  Unstructured volume raycasting in our prototype GPU renderer run locally
	  on six nodes, each with two GTX 1070s.}
	  \vspace{-1em}
  \end{figure}

\begin{figure}[!ht]
	\centering
	\includegraphics[width=0.8\columnwidth]{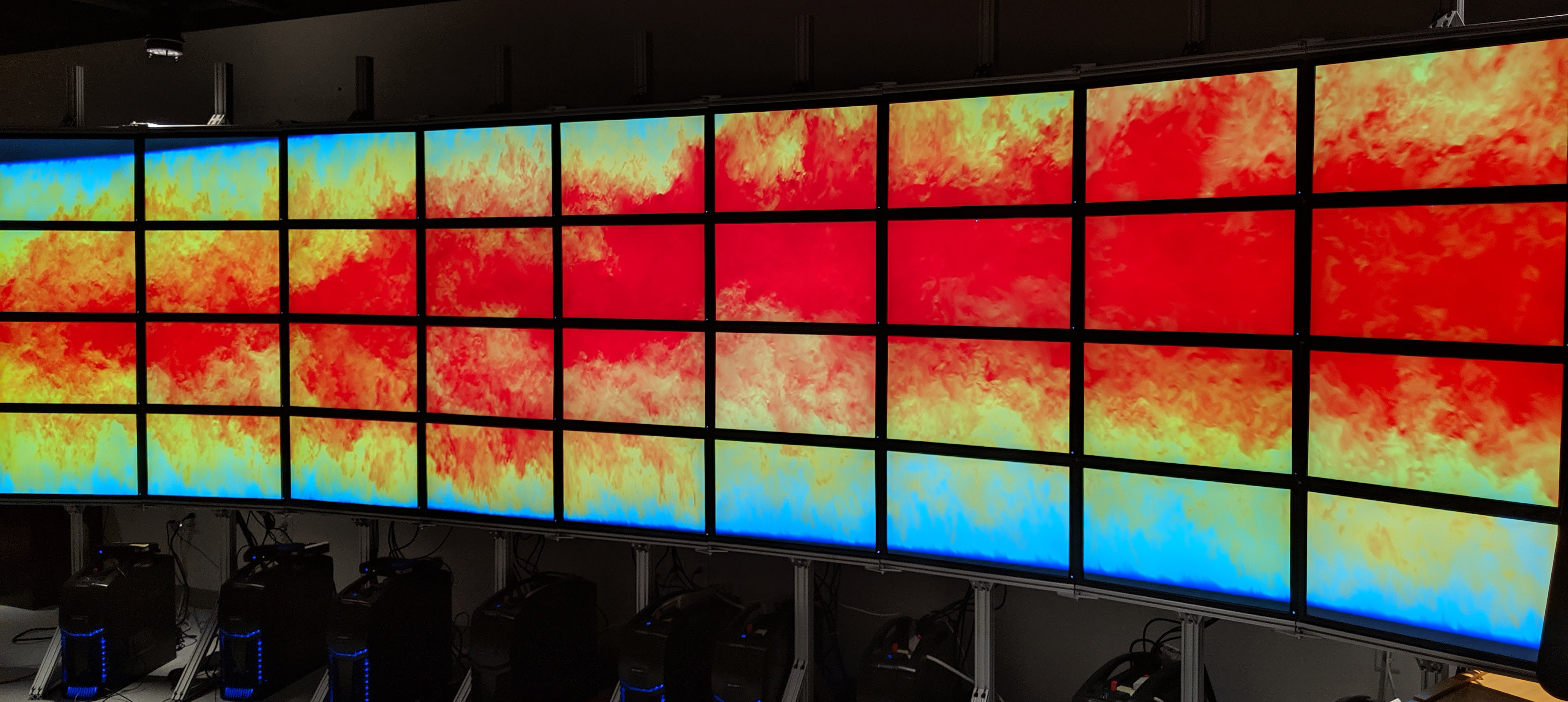}
	   \vspace{-0.5em}
	\caption{\label{fig:sci_stampede2_dns}%
		Data-parallel rendering of the 500GB DNS volume (10240$\times$7680$\times$1536 grid)
	  with OSPRay on 64 SKX nodes on Stampede2, streamed to the POWERwall in direct mode,
	  averaging 6-10~FPS.}
	  \vspace{-2.1em}
  \end{figure}

In~\Cref{fig:tile-size}, we evaluate the display performance when
using different tile sizes on the client. We find that small tile
sizes, which in turn require many small messages to be sent over the network,
underutilize the network and achieve poor performance.
Larger tile sizes correspond to larger messages,
reducing communication overhead and achieving better performance as a
result. This effect is more pronounced in the dispatcher mode, as
the overhead of the small tiles must be paid twice: once when sending
to the dispatcher, and again when sending from the dispatcher to the display.


In~\Cref{fig:compression}, we evaluate the performance impact of
the JPEG quality threshold set by the client. As display walls
are typically on the order of hundreds of mega-pixels, compression
is crucial to reducing the bandwidth needs of the system to
achieve interactive rendering performance.


In~\Cref{fig:scaling}, we evaluate the scalability of \emph{dw2} when increasing
the number of displays or clients. We find the direct mode scales well
with the number of displays and clients, since each client and display pair can communicate
independently, whereas the dispatcher mode introduces a bottleneck at the head node.

Based on the results of our parameter study,
we recommend using \emph{dw2} with a $128^2$ or $256^2$ tile size
with JPEG quality of 50-75, and we prefer the direct mode if the
underlying network architecture supports an all-to-all connection
between the clients and displays.

\subsection{Example Use Cases}
\label{sec:use_cases}

We demonstrate \emph{dw2} on interactive rendering of several medium- 
to large-scale datasets across the three display walls using a range of client hardware.
\Cref{fig:teaser,fig:sci_stampede2_dns} show medium- to large-scale
datasets rendered remotely on 64 or 128 Stampede2
Skylake Xeon nodes with OSPRay and streamed back to the POWERwall using the direct connection
mode.
In~\Cref{fig:sci_tetty_rtx}, we use our GPU prototype raycaster to
render across six nodes, each with two NVIDIA GTX 1070 GPUs, and displayed locally on the POWERwall
using the direct mode.
In~\Cref{fig:rattler_stampede2_moana}, we show the Moana Island Scene rendered
on Stampede2 with OSPRay and displayed locally on Rattler, using the dispatcher mode.
In~\Cref{fig:nucs} we render the Landing Gear AMR isosurface on-site using the eight node KNL
cluster and displayed on the NUCwall in direct mode.
For both on-site and remote rendering on CPU and GPU clusters,
\emph{dw2} allows renderers to achieve
interactive performance (also see the supplemental video).

%% file: summary.tex
\section{Discussion and Conclusion}

We have presented an open-source lightweight framework for rendering 
to large tiled display walls from a single source, based on a virtual frame buffer
abstraction concept. Our framework is easy to integrate into rendering
applications and provides the flexibility required to be deployed
across the display wall configurations typically found in
visualization centers. Moreover, we have demonstrated that combining
low-cost display nodes with remote rendering on an HPC resource can
be a compelling option for interactively driving tiled displays. 




%% file: acks.tex
\section*{Acknowledgements}
The authors wish to thank Jo\~{a}o Barbosa for helping running experiments on Rattler.
Additional support comes from the Intel Graphics and Visualization Institute
of XeLLENCE, the National Institute of General Medical Sciences of 
the National Institutes of Health under grant numbers P41 GM103545 and R24 GM136986, 
the Department of Energy under grant number DE-FE0031880,
NSF:OAC: Awards 1842042 and 1941085,
and NSC:CMMI: Award 1629660.
The authors thank the Texas Advanced
Computing Center (TACC) at The University of Texas at Austin for 
providing access to Rattler and Stampede2.